\documentclass[conference]{IEEEtran}
\pdfoutput=1
\usepackage{cite}
\usepackage{float}
\usepackage{tikz}
\usepackage{siunitx}
\usepackage{subcaption}
\usepackage{listings}
\usepackage{xparse}
\usepackage{xcolor}
\definecolor{codegreen}{rgb}{0,0.6,0}
\definecolor{codegray}{rgb}{0.5,0.5,0.5}
\definecolor{codepurple}{rgb}{0.58,0,0.82}
\definecolor{backcolour}{rgb}{0.95,0.95,0.92}
\lstdefinestyle{mystyle}{
    backgroundcolor=\color{backcolour},   
    commentstyle=\color{codegreen},
    keywordstyle=\color{magenta},
    numberstyle=\tiny\color{codegray},
    stringstyle=\color{codepurple},
    basicstyle=\ttfamily\footnotesize,
    breakatwhitespace=false,         
    breaklines=true,                 
    captionpos=b,                    
    keepspaces=true,                 
    numbers=left,                    
    numbersep=5pt,                  
    showspaces=false,                
    showstringspaces=false,
    showtabs=false,                  
    tabsize=2
}
\definecolor{scalebgcolor}{rgb}{0.08,0.52,0.80}

\usepackage[belowskip=0pt,aboveskip=0pt]{caption}
\usepackage{amsmath,amssymb,amsfonts}
\usepackage{algorithmic}
\usepackage{graphicx}
\usepackage{textcomp}
\usepackage{physics}

\def\BibTeX{{\rm B\kern-.05em{\sc i\kern-.025em b}\kern-.08em
    T\kern-.1667em\lower.7ex\hbox{E}\kern-.125emX}}
\begin{document}
\title{X-ray dark-field and phase retrieval without optics, via the Fokker--Planck equation}
\author{\Large Thomas A. Leatham,$^{1^{*}}$ \IEEEmembership{}David M. Paganin,$^{1}$ 
\IEEEmembership{}and Kaye S. Morgan$^{1}$ 
\IEEEmembership{} \\ \\
\normalsize $^{1}$ School of Physics and Astronomy, Monash University, Victoria, Australia \\
\normalsize  $^{*}$ thomas.leatham@monash.edu}

\maketitle
\begin{abstract}
Emerging methods of x-ray imaging that capture phase and dark-field effects are equipping medicine with complementary sensitivity to conventional radiography. These methods are being applied over a wide range of scales, from virtual histology to clinical chest imaging, and typically require the introduction of optics such as gratings. Here, we consider extracting x-ray phase and dark-field signals from bright-field images collected using nothing more than a coherent x-ray source and a detector. Our approach is based on the Fokker--Planck equation for paraxial imaging, which is the diffusive generalization of the transport-of-intensity equation. Specifically, we utilize the Fokker--Planck equation in the context of propagation-based phase-contrast imaging, where we show that two intensity images are sufficient for successful retrieval of both the projected thickness and the dark-field signal associated with the sample. We show the results of our algorithm using both a simulated dataset and an experimental dataset.  These demonstrate that the x-ray dark-field signal can be extracted from propagation-based images, and that sample thickness can be retrieved with better spatial resolution when dark-field effects are taken into account. We anticipate the proposed algorithm will be of benefit in biomedical imaging, industrial settings, and other non-invasive imaging applications.
\end{abstract}
\begin{IEEEkeywords}
x-ray imaging, phase retrieval, dark-field retrieval, propagation-based imaging, homogeneous samples, Fokker--Planck equation
\end{IEEEkeywords}
\section{Introduction}
X-ray imaging has been widely adopted in a range of fields, including medicine, security, and manufacturing industries, providing a way to probe the internal structure of a sample in a non-invasive manner. The traditional x-ray contrast mechanism is attenuation, where high-density objects reduce the intensity of the incident x-ray wavefield upon passing through the sample. In recent decades, phase-contrast imaging has been developed, where even low density/weakly-attenuating sample features are made visible, based on their alteration of the phase of an incident wavefield as it passes through the sample\cite{Wilkins2014evolution},\cite{Endrizzi}. Even more recently, sub-pixel features in samples have been rendered detectable by measuring an x-ray dark-field signal that results from the diffuse scattering of the incident wavefield from these sub-pixel features\cite{Paganin2020x}. The ability to detect small features with much-larger pixels means dark-field imaging has a significant dose-saving advantage, and it is already finding clinical application\cite{willer2021}.  Such a dark-field signal has been primarily captured using analyzer-based imaging\cite{Ando2001,Ando2004,Pagot2003} and grating interferometry\cite{Pfeiffer}. In this paper we exclusively focus on an x-ray method that has not been widely used for dark-field imaging, namely  propagation-based imaging (PBI) \cite{Cloetens1996,snigirev1995, Wilkins1996}, and propose a new phase and dark-field retrieval method. 
\par
Conventionally in PBI, dark-field effects have not been considered or have been assumed to be negligible. In this context, the transport-of-intensity equation (TIE)\cite{Teague1983} can be used to model the formation of x-ray intensity images at a detector located downstream of a sample, at a distance $z=\Delta$ along the optical axis. This free-space propagation results in bright/dark intensity fringes in the images \cite{Cloetens,snigirev1995}, which highlight changes in sample thickness or material, and which can be used in phase-retrieval algorithms to quantitatively recover sample information \cite{Paganin2002}.  The TIE describes coherent energy transport, i.e.~local conservation of the coherent optical energy flow of the transmitted x-ray beam as it propagates downstream of the sample. Recent work has shown that by accounting for the presence of unresolved microstructure (sub-pixel features) in the sample, PBI can incorporate a dark-field signal due to local sample-induced diffuse scatter \cite{Morgan2019,Paganin2019,Gureyev2020,Alaleh2022}. In doing so, the TIE is generalized by the x-ray Fokker--Planck equation, which for rotationally-invariant position-dependent small-angle x-ray scattering (SAXS) cones has the finite-difference form\cite{Morgan2019,Paganin2019}:
\begin{align}
\label{eqn:FPE}
I\qty(x,y,z=\Delta)\approx&I\qty(x,y,z=0)\\ \nonumber-&\frac{\Delta}{k}\grad_{\perp}\vdot\qty[I\qty(x,y,z)\grad_{\perp}\phi\qty(x,y,z)]_{z=0}\\ \nonumber
+&\Delta^{2}\laplacian_{\perp}\qty[D\qty(x,y)I\qty(x,y,z)]_{z=0}.
\end{align}
In the limit of zero diffusion, one recovers the TIE from the x-ray Fokker-Planck equation above. Here, \textit{$I\qty(x,y,z=\Delta)$} is the intensity of the x-ray wavefield recorded at a detector located at a propagation distance \textit{z=$\Delta$} downstream of the sample, \textit{$I\qty(x,y,z=0)$} is the intensity of the wavefield at the exit-surface (\textit{z=$0$}) of the sample, \textit{k} is the wavenumber of the wavefield corresponding to a wavelength \textit{$\lambda$} defined by $k=2\pi/\lambda$, \textit{$\phi\qty(x,y,z)$} is the phase of the wavefield, \textit{$D\qty(x,y)$} is the dimensionless diffusion coefficient describing local sample-induced SAXS and $\grad_{\perp}\equiv (\partial/\partial x,\partial/\partial y)$ is the gradient operator with respect to transverse coordinates $(x,y)$.

The first two terms on the right-hand side of $(\ref{eqn:FPE})$ comprise the finite-difference TIE part of the Fokker--Planck equation, and describe (i) the local attenuation of the incident wavefield by the sample being imaged\cite{Rontgen} (the $I\qty(x,y,z=0)$ term), (ii) the transverse shifting of the wavefield intensity due to the refractive effects induced by the sample\cite{Gureyev2006}, and (iii) the concentration/rarefaction of the intensity of the wavefield due to local focusing/defocusing effects\cite{Gureyev2006}. The final term on the right-hand side of $(\ref{eqn:FPE})$ comprises the diffusive part of the x-ray Fokker--Planck equation, and describes the position-dependent local blurring of the wavefield intensity due to the presence of unresolved microstructure within the volume of the sample. This local blurring of the wavefield intensity may be seen as a reduction in the visibility of a measured x-ray intensity distribution when captured at a detector located downstream of the sample, relative to the intensity that would be seen in the absence of unresolved microstructure. Accordingly, we associate the Fokker--Planck diffusion coefficient $D(x,y)$ with the dark-field signal due to unresolved sample microstructure. Here, visibility ($\textit{V}$) is defined using Michelson's definition\cite{michelson1995studies}:
\begin{equation}
    V=\frac{I_{\text{max}}-I_{\text{min}}}{I_{\text{max}}+I_{\text{min}}},
\end{equation}
where $I_{\text{max}}$ and $I_{\text{min}}$ are the maximum and minimum intensity values of the fringes in a given region of the recorded intensity pattern. Note that by keeping the $D\qty(x,y)$ term inside the transverse Laplacian in (\ref{eqn:FPE}), we are not assuming the diffusion coefficient to be spatially slowly-varying, as was assumed in \cite{Paganin2019}. The characterization given for the diffusion coefficient $D$ here differs to that given in \cite{Paganin2019}, $D_{\text{Paganin}}$, and in \cite{Morgan2019}, $D_{\text{Morgan}}$, through the relation 
\label{eqn:SAXScorrespondence}
$\Delta D\qty(x,y) =D_{\text{Morgan}}\qty(x,y,z=\Delta) =F D_{\text{Paganin}}\qty(x,y,z=\Delta).$
Due to the definition of the diffusion coefficient given in equation (40) of \cite{Paganin2019}, the least possible scattering due to SAXS is naturally restricted to zero, and hence the diffusion coefficient, as specified in this manuscript, must be manifestly non-negative, in order to describe a blurring effect rather than a focussing effect. 
\begin{figure}
\centering
\includegraphics[scale=0.145]{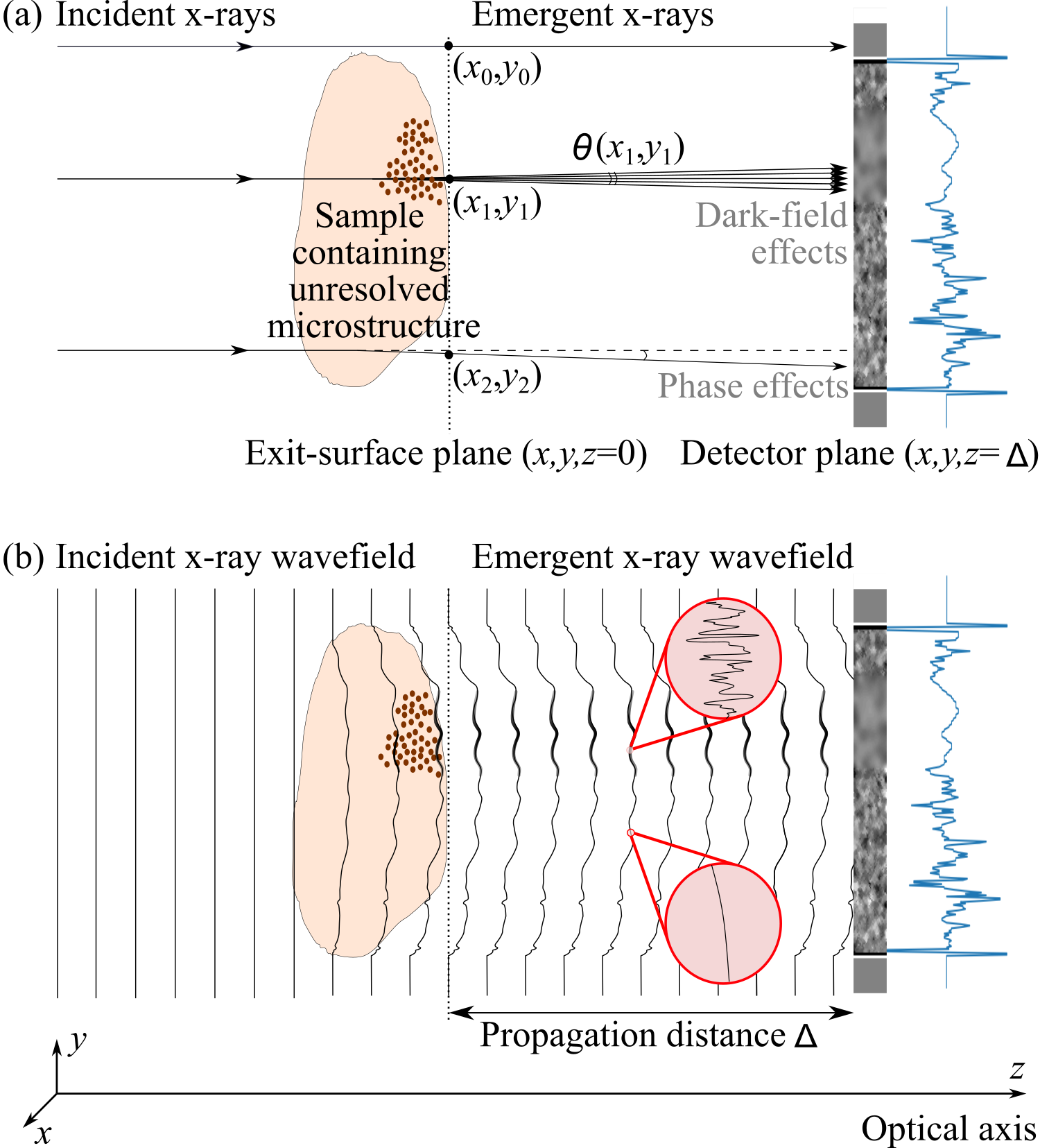}
\caption{Experimental setup to capture both phase and dark-field signals using propagation-based phase-contrast x-ray imaging. (a) In the ray picture incident x-rays are attenuated or transversely shifted by the sample. Additionally, unresolved microstructure present in the sample causes the emergent x-rays to be diffusely scattered, resulting in a SAXS cone. (b) In the wave picture, the incident x-ray wavefield is attenuated and acquires phase shifts as it passes through the sample. The phase of the exiting wavefield may be split into (i) a slowly-varying component associated with spatially resolved sample structure, giving rise to the propagation-based phase-contrast signal, and (ii) a rapidly-varying component associated with unresolved microstructure, giving rise to the dark-field signal.}
\label{fig:schematic}
\end{figure}

Consider the experimental setup shown in Fig.~\ref{fig:schematic}, where a sample containing unresolved microstructure is illuminated with an x-ray source (not shown), allowing for propagation-based intensity images to be captured at a detector located at a variable propagation distance $z=\Delta$ downstream of the sample. In the ray-optics description of this setup, as seen in Fig.~\ref{fig:schematic}(a), incident x-rays can pass outside the sample, or can interact with a region of the sample which may or may not contain unresolved microstructure. Rays which do not interact with the sample, such as the ray passing through the point $\qty(x_{0},y_{0})$ at the exit-surface plane, do not experience any change in intensity or phase. Hence, the intensity of these rays measured at the detector plane is the same as that which would have been measured in the absence of the sample. Rays which interact with a region of the sample that does not contain unresolved microstructure, such as the ray passing through the point $\qty(x_{2},y_{2})$, will experience attenuation and phase effects imparted by the sample, as described by the TIE part of the x-ray Fokker--Planck equation, i.e.~these rays have reduced intensity and are refracted by the sample. Rays which interact with a region of the sample that does contain unresolved microstructure, such as the ray passing through the point $\qty(x_{1},y_{1})$, will also experience both attenuation and refraction effects, but in addition, the regions of the images where such rays land will undergo a visibility reduction (dark-field effect) due to the unresolved microstructure contained in the sample. Upon passing through the sample, a fraction of these rays will be diffusely scattered by the unresolved microstructure, resulting in a spray of emergent x-rays through an opening angle $\theta$, taking the shape of a cone, landing at the detector plane with width proportional to the propagation distance $\Delta$. It is this spray of emergent x-rays that gives rise to the local visibility reduction seen in the intensity pattern recorded at the detector, a characteristic hallmark of the dark-field signal.  We can also explain these effects with reference to a wave-optics description, as seen in Fig.~\ref{fig:schematic}(b). Upon interacting with the sample and subsequently exiting the sample, the incident x-ray wavefield acquires a change in phase. The presence of unresolved microstructure in the sample causes the phase of the incident wavefield to split into two components: a slowly-varying component, corresponding to smooth (resolvable) sample features, and a rapidly-varying component, corresponding to fine (unresolved) sample features \cite{Nesterets}. The slowly-varying phase component (unbolded sections of emergent wavefield -- the lower red inset shows these slow phase variations) is what we retrieve as the phase signal. The rapidly-varying phase component, along with the self-interference of the emergent wavefield, causes the diffusion of the intensity variations of the emergent wavefield as it propagates to the detector plane (bolded sections of emergent wavefield -- the upper red inset shows these rapid phase variations), resulting in regions of reduced visibility in the recorded intensity pattern. This blurring effect, described in Nesterets \textit{et al.}\cite{Nesterets}, is also apparent in biological samples, for example in images of mice lungs taken with a large sample-to-detector propagation distance \cite{Yagi}.
Hence, it is this rapidly-varying phase component that gives rise to the dark-field signal in the wave-optics description. Key to the retrieval method presented in this paper is the fact that the PBI phase fringes evolve differently with propagation  to the dark-field diffusive effects, as described by the Fokker--Planck equation (\ref{eqn:FPE}), and hence the two can be separated.

The Fokker--Planck equation describes the transverse redistribution of optical energy carried by the incident wavefield,  as it propagates downstream of the sample. On account of the conservation of energy, if a fraction of the wavefield is diffusely scattered by the sample, the remaining fraction of the wavefield must be coherently transported. Hence, there is a bifurcation of the optical flow of the incident wavefield into the coherent and diffusive energy channels. The Fokker--Planck equation may thus be viewed as the natural diffusive generalization of the TIE, which simultaneously models attenuation, phase and dark-field effects in PBI settings. The Fokker--Planck equation has been applied to grid/grating-based imaging in the context of the forward problem as seen in Morgan \& Paganin\cite{Morgan2019}, and has also been used to retrieve phase and dark-field signals in the context of the inverse problem applied to x-ray speckle-tracking\cite{MIST},\cite{MISTdirectional}. Both of these dark-field imaging techniques, as well as analyzer-based dark-field imaging, have the disadvantage of requiring extra hardware in the form of optical elements to extract phase and dark-field contrast. Through the PBI model provided by the Fokker--Planck equation, we derive an algorithm to perform simultaneous x-ray phase and dark-field retrieval without optics. Our method uses virtual-optics software\cite{PaganinOmni2004} rather than optical hardware, to extract a dark-field signal from bright-field data.  This software encodes our closed-form analytical solution to the Fokker--Planck formulation of propagation-based image formation in the presence of unresolved sample microstructure. It is the goal of this paper to outline how this retrieval is possible with homogeneous samples and to demonstrate the results of our algorithm with both simulated and experimental datasets. One should keep in mind that when the phrase `no optics' is referred to within this manuscript, what is really meant is that no optical elements are used in the experimental setup apart from the source, sample and detector.

The structure of this paper is as follows. Section II derives our PBI phase and dark-field retrieval algorithm and provides a corresponding physical interpretation. Section III demonstrates the use of our algorithm with a simulated dataset, and section IV shows the results of applying our algorithm to an experimental dataset. Section V discusses the broader implications of this work and section VI outlines possible directions for future research and also provides some concluding remarks. 

\section{Derivation of the PBI phase and dark-field retrieval method}

Assume that a thin, static, non-crystalline, and non-magnetic sample is illuminated by quasi-monochromatic $z$-directed x-ray plane waves of incident intensity \textit{$I_{0}$}. Further assume that all polarization-sensitive effects can be ignored. Provided the Fresnel number \cite{SalehTeichBook} is much larger than unity, our starting point is the x-ray Fokker--Planck equation $\qty(\ref{eqn:FPE})$. We specialize to the scenario of a single-material sample located immediately upstream of the plane $z=0$, with a projected thickness $T\qty(x,y)$ along the $z$ direction, as follows. Invoking the projection approximation\cite{Coherent}, the phase of the wavefield at the exit-surface of the sample and the exit-surface intensity of the wavefield are given by:
\begin{equation}
\label{eqn:phaseproj}
\phi\qty(x,y,z=0)=-k\delta T\qty(x,y)
\end{equation}
and
\begin{equation}
\label{eqn:intensityproj}
I\qty(x,y,z=0)=I_{0}\exp\qty[-\mu T\qty(x,y)]
\end{equation}
respectively. Here \textit{$\delta$} is the real decrement of the complex refractive index of the sample 
\begin{equation}
\label{eqn:index}
n\qty(x,y,z)=1-\delta\qty(x,y,z)+i\beta\qty(x,y,z)
\end{equation}
and the linear attenuation coefficient  \textit{$\mu$} is related to \textit{$\beta$} via 
\begin{equation}
\label{eqn:mu}
\mu=2k\beta.
\end{equation}
Inserting (\ref{eqn:phaseproj}) and (\ref{eqn:intensityproj}) into (\ref{eqn:FPE}) yields 
\begin{align}
\label{eqn:FPESM}
I\qty(x,y,z=\Delta)&=I_{0}\exp\qty[-\mu T\qty(x,y)]\\ \nonumber-&\frac{\Delta I_{0}}{k}\grad_{\perp}\vdot\qty[\exp\qty[-\mu T\qty(x,y)]\grad_{\perp}\qty(-k\delta T\qty(x,y))]\\ \nonumber
+&\Delta^{2} I_{0}\laplacian_{\perp}\qty[D\qty(x,y)\exp\qty[-\mu T\qty(x,y)]].
\end{align}
We now employ the identity\cite{Paganin2002}
\begin{align}
\label{eqn:NoPrism}
\grad_{\perp}\vdot\qty[\exp\qty[-\mu T\qty(x,y)]\grad_{\perp}\qty(-k\delta T\qty(x,y))]\\ \nonumber=\frac{k\delta}{\mu}\laplacian_{\perp}\exp\qty[-\mu T\qty(x,y)]
\end{align}
and make use of the fact that the single-material assumption implies the ratio \textit{$\delta\qty(x,y,z)/\beta\qty(x,y,z)$} to be the same at all locations within the sample.  Hence, (\ref{eqn:FPESM}) becomes: 
\begin{align}
\label{eqn:FPESM2}
I\qty(x,y,z=\Delta)&=I_{0}\exp\qty[-\mu T\qty(x,y)]\\ \nonumber-&\frac{\Delta I_{0}\delta}{\mu}\laplacian_{\perp}\exp\qty[-\mu T\qty(x,y)]\\ \nonumber
+&\Delta^{2} I_{0}\laplacian_{\perp}\qty[D\qty(x,y)\exp\qty[-\mu T\qty(x,y)]].
\end{align}

We note that (\ref{eqn:FPESM2}) contains two unknown quantities to solve for, namely \textit{$D\qty(x,y)$} and \textit{$T\qty(x,y)$}. Two equations or measurements are hence required to find these two quantities. The simplest variable to change is the propagation distance \textit{$\Delta$}, although it would also be possible to write (\ref{eqn:FPESM2}) for two different energies, since \textit{$\delta$} and \textit{$\mu$} depend on the energy of the x-ray beam. Note that we are assuming the plane wave approximation so that what is classified as either resolved or unresolved does not change with propagation distance. Here we take the approach of changing the propagation distance, and so to proceed we write the \textit{$z=\Delta_{1}$} and \textit{$z=\Delta_{2}$} cases of (\ref{eqn:FPESM2}), eliminating the \textit{$D\qty(x,y)$} term:
\begin{align}
\label{eqn:thickop}
\qty(1 - \frac{\delta}{\mu}\frac{\Delta_{1}\Delta_{2}}{\Delta_{1}+\Delta_{2}}\laplacian_{\perp})\exp\qty[-\mu T\qty(x,y)] \quad\quad\quad\quad\quad \\ \nonumber = \frac{\Delta_{2}^{2}I\qty(x,y,z=\Delta_{1})-\Delta_{1}^{2}I\qty(x,y,z=0;\Delta_{2})}{I_{0}\qty(\Delta_{2}^{2}-\Delta_{1}^{2})}.
\end{align}
This has the same form as seen in the derivation of a routinely used homogeneous-object TIE phase-retrieval algorithm\cite{Paganin2002}, and can hence be solved using the same Fourier-transform method. This gives (cf. equation (62) in \cite{Paganin2019}):
\begin{align}
\label{eqn:thickness}
T&\qty(x,y)=\frac{-1}{\mu}\times\\ \nonumber \ln&\qty{\!\mathcal{F}^{-1}\!\qty[\frac{\mathcal{F}\qty[\Delta_{2}^{2}I\qty(x,y,z=\Delta_{1})-\Delta_{1}^{2}I\qty(x,y,z=\Delta_{2})]}{I_0\qty(\Delta_{2}^{2}-\Delta_{1}^{2}+\frac{\delta}{\mu}\Delta_{1}\Delta_{2}\qty(\Delta_{2}-\Delta_{1})\qty(k_{x}^{2}+k_{y}^{2}))\!}]}.
\end{align}
Above, \textit{$\mathcal{F}$} denotes Fourier transformation with respect to \textit{$x$} and \textit{$y$}, the corresponding Fourier-space coordinates are denoted by \textit{$(k_x,k_y)$}, \textit{$\mathcal{F}^{-1}$} denotes inverse Fourier transformation with respect to \textit{$k_x$} and \textit{$k_y$}, and we use the Fourier transform convention found in \cite{Paganin2002} and \cite{Coherent}.

With \textit{$T\qty(x,y)$} given by (\ref{eqn:thickness}), the \textit{$z=\Delta_{1}$} version of (\ref{eqn:FPESM2}) can be rearranged into the form:
\begin{align}
\label{eqn:Dlap}
&\laplacian_{\perp}\qty[D\qty(x,y)\exp\qty[-\mu T\qty(x,y)]]\\ &\nonumber=\frac{I\qty(x,y,z=\Delta_{1})}{I_{0}\Delta_{1}^{2}}
-\qty(\frac{1}{\Delta_{1}^{2}}-\frac{\delta}{\mu \Delta_{1}}\laplacian_{\perp})\exp\qty[-\mu T\qty(x,y)].
\end{align}
Solving this Poisson equation for the dark-field signal, we obtain 
\begin{align}
\label{eqn:DF}
&D\qty(x,y)=\exp\qty[\mu T\qty(x,y)]\times\\&\nonumber\grad_{\perp}^{-2} \qty[\frac{I\qty(x,y,z=\Delta_{1})}{I_{0}\Delta_{1}^{2}}
-\qty(\frac{1}{\Delta_{1}^{2}}-\frac{\delta}{\mu \Delta_{1}}\laplacian_{\perp})\exp\qty[-\mu T\qty(x,y)]],
\end{align}
where the inverse Laplacian is defined as a pseudo-differential operator by \cite{Gureyev1997RapidQP} 
\begin{equation}
\label{eqn:invlap}
\grad_{\perp}^{-2}=-\mathcal{F}^{-1}\frac{1}{k_{x}^{2}+k_{y}^{2}}\mathcal{F}. 
\end{equation}
We note that the previous expression for the inverse Laplacian is singular when \textit{$\qty(k_{x},k_{y})=\qty(0,0)$}, so to avoid this singularity in computations we make the replacement 
\begin{equation}
\label{eqn:replace}
\frac{1}{k_{x}^{2}+k_{y}^{2}}\rightarrow \frac{1}{k_{x}^{2}+k_{y}^{2}+\varepsilon}
\end{equation}
where \textit{$\varepsilon > 0$} is small compared to \textit{$k_{x}^{2}+k_{y}^{2}$} (except for the vicinity of the  origin of Fourier space). The replacement (\ref{eqn:replace}) regularizes the inverse transfer function \textit{$H(k_{x},k_{y})=1/(k_{x}^{2}+k_{y}^{2})$} by replacing the blow-up at the origin of Fourier space with a finite non-zero DC term \textit{$1/\varepsilon$}, for fixed \textit{$\varepsilon$}. 

To apply the retrieval method derived in this section, we need to first capture two intensity images at two different propagation distances, and insert these intensity images and the relevant parameters into (\ref{eqn:thickness}) to retrieve the thickness of the sample. We then need to use the retrieved thickness image and one of the intensity images to reconstruct the dark-field signal according to (\ref{eqn:DF}), where the inverse Laplacian is computed according to (\ref{eqn:invlap}) and (\ref{eqn:replace}). It is also worth noting that we can reconstruct the dark-field signal using a third intensity image taken at a propagation distance that is different from the two distances used to reconstruct the projected thickness of the sample, provided that this reconstruction of the projected thickness is relatively stable with respect to the propagation distances used to perform the reconstruction. In fact, so long as we obey the simple rule of thumb that longer propagation distances are beneficial to render dark-field effects visible and shorter propagation distances are beneficial to achieve high-spatial-resolution phase effects and hence thickness retrieval, the reconstruction process for both the projected thickness and dark-field signals will work well. An example showing the visible increase in dark-field effects with distance can be found in the movie in the Supplementary Materials II, where a sequence of propagation-based images captured at distances ranging from \SI{0.5}{\meter} to \SI{7}{\meter} of a small plastic tube filled with agarose powder, attached to the centre of a green seed pod from a {\em Liquidambar styraciflua} tree using Kapton tape (see Appendix IV). The dark-field-associated-blurring in the center of the images, due to the powder, becomes first most visible at a distance of \SI{3}{\meter}, indicating that distances around this value are beneficial for both phase and dark-field retrieval for this sample. This rule of thumb indicates that there is an inherent trade-off between visualizing phase and dark-field effects in the Fokker--Planck description; the propagation distances chosen should be large enough to clearly render dark-field effects, but should not be so large that phase effects are hard to visualize and detail is lost. 
The method is designed for the case that all propagation distances are chosen are such that the near-field condition is satisfied, i.e.~the Fresnel number \cite{SalehTeichBook} should be much greater than unity for the spatially-resolved projected sample structure. More specifically, this condition can be formulated as
\begin{equation}
\label{eqn:Fresnel}
N_{\text{F}}=\frac{a^{2}}{\lambda \Delta} \gg 1,
\end{equation}
where $N_{\text{F}}$ denotes the Fresnel number, $a$ is the smallest spatially-resolved feature size present, $\lambda$ is the x-ray wavelength and $\Delta$ is the propagation distance. One should note that the value of $a$ cannot be smaller than twice the detector pixel size\cite{weitkamp}. Hence, a lower bound for the Fresnel number for a given x-ray energy and given propagation distance is 
\begin{equation}
\label{eqn:Fresnelmin}
N_{\text{F,min}}=\frac{4d^{2}}{\lambda \Delta},
\end{equation}
where $d$ is the detector pixel size.
\begin{figure}[ht]
\centering
\hspace*{-1.5em}
\includegraphics[width=1.1\linewidth]{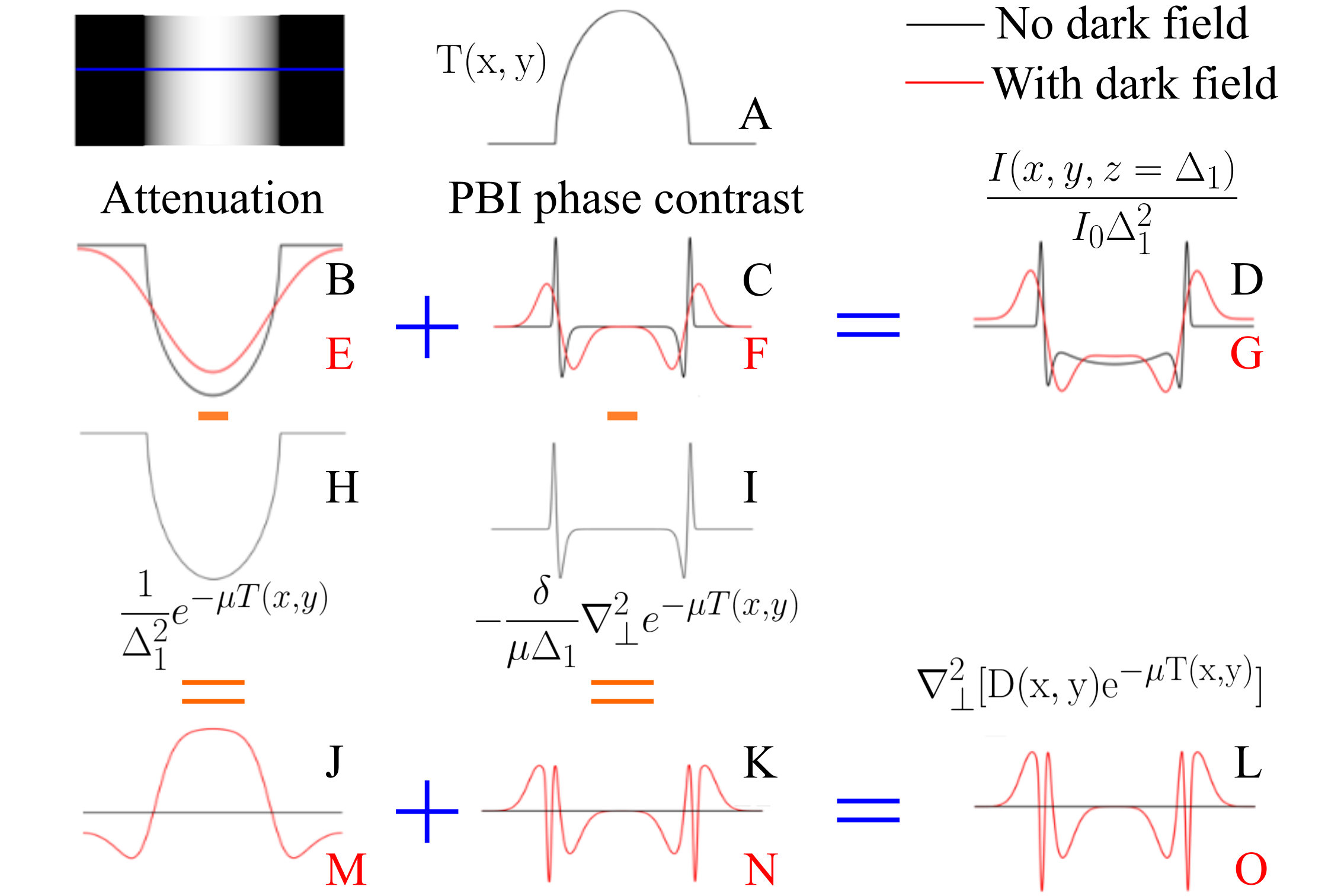}
\caption{Attenuation and PBI phase-contrast profiles of a sample for the projected thickness of the sample and each term in the bottom line of (\ref{eqn:Dlap}), where profiles in black correspond to zero dark-field and profiles in red correspond to a non-zero dark-field. When the profiles for each term on the bottom line of (\ref{eqn:Dlap}) are added together, we obtain the contributions sourced from local blurring of attenuation and phase contrast to the \textit{$\laplacian_{\perp}\qty[D\qty(x,y)\exp\qty[-\mu T\qty(x,y)]]$} term. Note that the orange `-' and `=' signs give column-wise relationships while the blue `+' and `=' signs give row-wise relationships.}
\label{fig:profiles}
\end{figure}

In closing this section, we provide an interpretation of (\ref{eqn:Dlap}) in terms of attenuation and PBI phase contrast intensity profiles for both the absence and presence of a dark-field signal, as shown in Fig.~\ref{fig:profiles}. In terms within the Fokker-Planck solution (\ref{eqn:DF}), the retrieved thickness is effectively used to predict both attenuation and phase effects in the absence of any dark-field signal. This prediction is then compared to the observed intensity image, where any differences are interpreted as resulting from dark-field effects. To illustrate this, we consider the case of imaging a cylinder as shown in Fig.~\ref{fig:profiles}, taking a thickness profile across the cylinder, such as the blue line, resulting in trace `A'.  The profiles for the exponential term in the final line of (\ref{eqn:Dlap}) take a similar shape, describing  attenuation expected by the sample, and the Laplacian term predicts phase effects for a sample of this given thickness. The profile for the scaled intensity image in the bottom line of (\ref{eqn:Dlap}) can have different shapes corresponding to the cases of zero dark-field (black profiles) and a non-zero dark-field signal (red profiles). When the dark-field signal is exactly zero (i.e.~no unresolved microstructure or edge-scattering effects), we would expect to see a profile such as trace `D' for the scaled intensity image, which can be broken into the attenuation profile trace `B' and phase-contrast profile trace `C'. In this case of zero dark-field, the attenuation profile (trace `B') is identical to that of the exponential term (trace `H'), and the phase-contrast profile (trace `C') is identical to that of the Laplacian term (trace `I'). We emphasize that we must be in the near-field regime to match the Laplacian and phase-contrast edge effects, and the propagation distance should be chosen to satisfy this condition, as per (\ref{eqn:Fresnel}) and (\ref{eqn:Fresnelmin}). For a non-zero dark-field signal, we would expect to see a profile such as trace `G' for the scaled intensity image. Trace `G' can be broken into the attenuation (trace `E') and phase contrast profiles (trace `F'), which  are locally blurred or diffused versions of traces `B' and `C', respectively. The subtraction of the profiles for the three terms in the final line of (\ref{eqn:Dlap}), which will be equal to the first line, \textit{$\laplacian_{\perp}\qty[D\qty(x,y)\exp\qty[-\mu T\qty(x,y)]]$}, is shown as traces `J' and `K' for the case of zero dark-field, and traces `M' and `N' for the case of non-zero dark-field. For zero dark-field, the attenuation and phase contrast contributions to this term sum together with the observed image to give zeros everywhere (trace `L'), resulting in a recovered dark-field signal that is identically zero everywhere. In the case of a non-zero dark-field signal, traces `M' and `N' sum together to give trace `O', which is not flat. Hence, when the inverse Laplacian is applied and the $\exp\qty[-\mu T\qty(x,y)]$ term is divided out, the dark-field signal recovered will not be zero. This observation indicates that the dark-field signal will be seen from a blurring of local intensity oscillations at the detector, whether those intensity oscillations are created by sample attenuation or sample-induced phase shifts, described by each of the terms in the final line of (\ref{eqn:Dlap}). We may also think of the dark-field signal as an obstruction which measures the extent to which the intensity measured at a detector fails to be described by the transport-of-intensity equation \cite{Gureyev2020}.

\section{Simulated PBI data}

\begin{figure}[ht]
\centering 
\includegraphics[width=\linewidth]{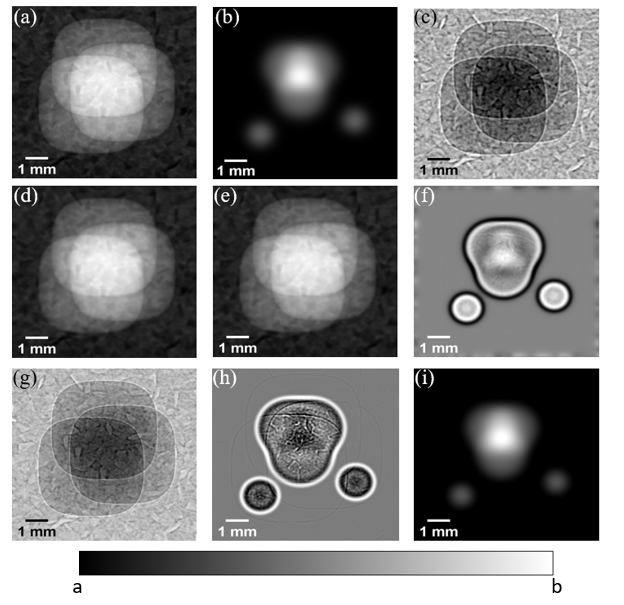}
\vspace*{0.3em}
\caption{All images here are shown with a linear grayscale, where $a$ is the minimum value and $b$ is the maximum value: (a) Simulated sample thickness image \textit{$T(x,y)$} of PMMA squircles with a piece of sandpaper providing background thickness variations ($a=\SI{8e-5}{\meter} $, $b=\SI{120e-5}{\meter}$). (b) Simulated dark-field image \textit{$D(x,y)$} of heavily smoothed PMMA spheres ($a=0$, $b=3\times 10^{-11}$). (c) Scaled intensity image corresponding to the \textit{$I\qty(x,y,z=\Delta_{1})/(I_{0}\Delta_{1}^{2})$} term of (\ref{eqn:Dlap}) at a propagation distance of \SI{65}{\centi\meter} ($a=\SI{2.31}{\meter^{-2}}$, $b=\SI{2.38}{\meter^{-2}}$). (d) Retrieved sample thickness using the TIE-based method of \protect\cite{Paganin2002} with the \SI{30}{\centi\meter} intensity image ($a=\SI{8e-5}{\meter}$, $b=\SI{120e-5}{\meter}$). (e) Retrieved sample thickness using (\ref{eqn:thickness}) in this paper with the \SI{30}{\centi\meter} and \SI{50}{\centi\meter} intensity images ($a=\SI{8e-5}{\meter}$, $b=\SI{120e-5}{\meter}$). (f) Difference between panels d) and e) ($a=\SI{-159e-7}{\meter}$, $b=\SI{128e-7}{\meter}$). (g) Difference between the exponential and Laplacian terms given in the latter half of the second line of (\ref{eqn:Dlap}) during dark-field retrieval ($a=\SI{2.31}{\meter^{-2}}$, $b=\SI{2.38}{\meter^{-2}}$). (h) Difference of c) and g), corresponding to \textit{$\laplacian_{\perp}\qty[D\qty(x,y)\exp\qty[-\mu T\qty(x,y)]]$} in the dark-field retrieval via equation (\ref{eqn:Dlap}) ($a=\SI{-688e-5}{\meter^{-2}}$, $b=\SI{477e-5}{\meter^{-2}}$). (i) Retrieved dark-field signal using (\ref{eqn:DF}) with the \SI{65}{\centi\meter} intensity image ($a=0$, $b=13\times 10^{-10}$).}
\label{fig:Sim}
\end{figure}

We now test the retrieval method using a simulated sample of three overlapping `squircles', each of dimensions $400$ pixels $\times$ $400$ pixels (\SI{4.6}{\milli\meter} $\times$ \SI{4.6}{\milli\meter}), embedded in an array of size $700$ pixels $\times$ $700$ pixels (\SI{8.05}{\milli\meter} $\times$ \SI{8.05}{\milli\meter}) with a sandpaper thickness image (retrieved by TIE from experimental data) providing more natural thickness variations across the entire array than seen with the perfectly-smooth squircles alone (see Fig.~\ref{fig:Sim}(a) for the resulting thickness map). We set the squircles to be made from PMMA and the energy of the x-ray beam to be  \SI{25}{\kilo\electronvolt}. The delta and beta values for PMMA at this energy are \textit{$\delta=\num{4.27e-07}$}
and $\beta=\num{7.00e-11}$\cite{Chantler}, and we set the pixel size to be \SI{11.5}{\micro\meter}. A completely independent dark-field signal was simulated using a combination of larger spheres of radius $100$ pixels and smaller spheres of radius $50$ pixels, with the dark-field signal strength specified in terms of a blurring width, with this blur width proportional to the thickness of the dark-field spheres. The thickness distribution of these dark-field spheres was also smoothed using a two-dimensional Gaussian function of $20$ pixels in standard deviation, to soften the edges of the spheres, thereby simulating a more realistic dark-field signal which would be seen in experiment (see Fig.~\ref{fig:Sim}(b) for the resulting dark-field map). No attenuation or phase effects are associated with these spheres, to be confident the recovered dark-field signal is sourced from dark-field effects alone. 

To perform our simulation we first calculated the wavefield at the exit-surface of the sample using the projection approximation and Beer's law to calculate the phase and intensity, respectively. We then propagated over the simulated sample-to-detector distances of \SI{30}{\centi\meter}, \SI{50}{\centi\meter} and \SI{65}{\centi\meter}, using the two-Fourier-transform representation of the Fresnel propagator \cite{Coherent}, to provide three simulated propagation-based intensity images. We then used the simulated dark-field signal to determine how much to locally blur these intensity images, by locally spreading the intensity at each detector pixel to the surrounding pixels, with a Gaussian full-width half-maximum (FWHM) of position-dependent width proportional to the strength of the simulated dark-field signal at that transverse ($x,y$) location within the sample. A two-dimensional Gaussian function of position-independent standard deviation equal to $2$ pixels in both the $x$ and $y$ directions was then used to further blur the intensity images, simulating blurring by a point-spread function (PSF) associated with a typical detector. A mathematical description of our simulation model can be found in Appendix I. These images were then taken as inputs to our retrieval method, and the projected thickness and dark-field signal were recovered according to (\ref{eqn:thickness}) and (\ref{eqn:DF}). One should be aware that the method described in this paragraph is not the only method available for incorporating the effect of local scattering. For example, one could use a scalar wave equation\cite{WolfPolar} or complex transmission function formalism\cite{Gureyev2020} to incorporate such scattering.

The sample projected thickness, recovered using (\ref{eqn:thickness}) and given in Fig.~\ref{fig:Sim}(e), and the dark-field signal, recovered using (\ref{eqn:DF}) and shown in Fig.~\ref{fig:Sim}(i), are consistent with the simulated sample. While the projected thicknesses recovered using the single-image method of Paganin \textit{et al.}~(Fig.~\ref{fig:Sim}(d)) and the dual-image Fokker--Planck method (Fig.~\ref{fig:Sim}(e)) appear identical, there is in fact a qualitative and quantitative difference between these images. This difference is due to the fact that the Fokker--Planck method explicitly takes into account the dark-field signal present in the simulated intensity image shown in Fig.~\ref{fig:Sim}(c). This difference is highlighted in Fig.~\ref{fig:Sim}(f), which shows the result of subtracting Fig.~\ref{fig:Sim}(d) from Fig.~\ref{fig:Sim}(e). As can be seen from Fig.~\ref{fig:Sim}(f), the difference between the projected thicknesses using the two different methods is related to the strength of the dark-field signal present in the reconstruction process. In regions where this dark-field signal is zero, i.e.~outside the dark-field-generating spheres, the difference between the projected thicknesses is smallest, while this difference is strongest in regions where the dark-field signal is strongest or where the dark-field signal changes quickly (see Fig.~4 of Morgan \textit{et al.}\cite{Morgan2019}), besides the effects seen at the borders of the image. See Appendix II for quantitative measures of the accuracy of the thickness reconstructions provided by both the TIE method of \cite{Paganin2002} and our Fokker-Planck method for the various possible propagation distances listed in this section.

The other key result of this section is the ability of our method to reconstruct the dark-field signal present in the simulated data, as can be seen by comparing Fig.~\ref{fig:Sim}(i) to Fig.~\ref{fig:Sim}(b). Our reconstruction method detects both those objects that generate a strong dark-field (the three larger overlapping spheres), and those that generate a weak dark-field (the smaller isolated spheres). Due to the detector point-spread function blurring, the simulated and retrieved dark-field signals have slightly different numerical values. Quantitative measures of the various reconstructed dark-field signals at the different propagation distances used in our simulations can be found in Appendix II. Additionally, the effect of noise in our simulations in terms of the qualitative and quantitative nature of the reconstruction process is discussed in Appendix III.

\section{Experimental PBI data}

Encouraged by the results of our simulation, we turn to demonstrating how this approach can be applied to experimental data. To do this, we collected propagation-based intensity images at the Australian Synchrotron on the Imaging and Medical Beamline (IMBL) in Hutch 3B.  The sample consisted of polystyrene microspheres, \SI{1}{\micro\meter} in diameter, contained in a sample tube made from PMMA. The sample was placed on a dedicated table located approximately \SI{130}{\meter} from the source, where x-ray photons of energy \SI{30}{\kilo\electronvolt} were produced by synchrotron radiation from a \SI{2}{\tesla} dipole bending magnet. At this energy, PMMA has $\delta=\num{2.96e-07}$
and $\mu=\SI{36.1}{\meter^{-1}}$, and we used these values to represent the complex refractive index of the sample. Using the PMMA values to represent the whole sample here is a good approximation since there is a relatively small difference between the $\delta$ and $\mu$ values for PMMA and polystyrene at an energy of $\SI{30}{\kilo\electronvolt}$ (approximately $11.9\%$ for the $\delta$ values and approximately $25.3\%$ for the $\mu$ values\cite{Chantler}).  The detector used to image the sample had an effective pixel size of \SI{18}{\micro\meter}, and by placing the detector at distances of \SI{1}{\meter}, \SI{2}{\meter}, and \SI{3}{\meter} downstream of the sample, we captured propagation-based intensity images exhibiting both phase and dark-field effects at each of these propagation distances. The exposure time for each image was \SI{1}{\second}.  Thirty exposures were captured at each propagation distance, and averaged before flat-field and dark-field correction. The flat-field and dark-field corrected images were then resized to account for slight magnification differences using the source-to-sample distance and finally translated and registered to each other to sub-pixel accuracy in order to mitigate alignment artifacts. All image processing and data analysis was done using Python3 code on a desktop machine. In particular, the translation and registration of the images was achieved using the `phase\_cross\_correlation' function from the registration module of the freely accessible scikit-image library\cite{scipy}, with an upsampling factor of 1000 as the input to this function. Furthermore, any apparent truncation of the sample was handled by mirroring the relevant images whenever using Fourier transforms in order to enforce the necessary periodicity conditions and hence avoid cross-talk between opposite borders of the images. 
\par
\begin{figure*}[ht]
\centering 
\includegraphics[width=\linewidth]{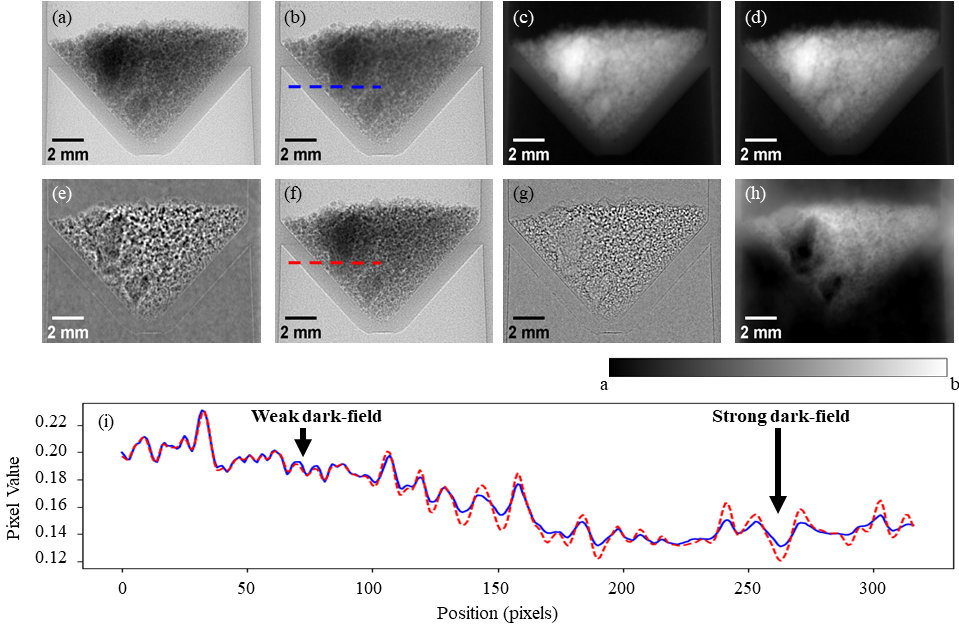}
\caption{(a) Propagation-based intensity image of the polystyrene microspheres contained in the PMMA sample tube captured at a propagation distance of \SI{1}{\meter} ($a=0.36$, $b=0.96$). (b) Scaled propagation-based intensity image (first term in the last line of (\ref{eqn:Dlap})) of the polystyrene microspheres taken at a propagation distance of \SI{2}{\meter} ($a=\SI{0.08}{\meter^{-2}}$, $b=\SI{0.25}{\meter^{-2}}$). (c) Retrieved thickness using TIE method with intensity image taken at \SI{1}{\meter} propagation ($a=\SI{40e-4}{\meter}$, $b=\SI{258e-4}{\meter}$). (d) Retrieved thickness using our Fokker--Planck method with intensity images taken at \SI{1}{\meter} and \SI{2}{\meter} propagation ($a=\SI{40e-4}{\meter}$, $b=\SI{258e-4}{\meter}$). (e) Difference between (c) and (d) ($a=\SI{-5e-4}{\meter}$, $b=\SI{5e-4}{\meter}$). (f) Dark-field-free intensity estimation at \SI{2}{\meter} according to the second term in the bottom line of (\ref{eqn:Dlap}) ($a=\SI{0.08}{\meter^{-2}}$, $b=\SI{0.25}{\meter^{-2}}$). (g) The Laplacian of the product of the dark-field signal and the $\exp[-\mu T\qty(x,y)]$ term, corresponding to (\ref{eqn:Dlap}) ($a=\SI{-0.06}{\meter^{-2}}$, $b=\SI{0.04}{\meter^{-2}}$). (h) Dark-field signal found using (\ref{eqn:DF}) ($a=0$, $b=11\times 10^{-10}$). (i) Profiles taken along blue and red dashed lines in (b) and (f), showing differences in visibility between the two terms on the last line of (\ref{eqn:Dlap}).}
\label{fig:Experimental}
\end{figure*}

See the movie in Supplementary Materials I for the full sequence of the propagation-based images collected, noting a reduction in local contrast with increasing propagation distance in the center of the image, where the greatest number of polystyrene microspheres are seen in projection.  Fig.~\ref{fig:Experimental} shows the results of using the \SI{1}{\meter} and \SI{2}{\meter} (Fig.~\ref{fig:Experimental}(a)) propagation-based intensity images to recover the projected thickness and dark-field images of the sample. These propagation distances were used as they provided a balanced trade-off between visualizing phase and dark-field effects, i.e.~the projected thickness and dark-field signal can be recovered accurately using this pair of propagation distances. As with our simulations in section III, we see subtle differences between the sample thickness retrieved using our Fokker-Planck approach (Fig.~4(d)), and the TIE approach of \cite{Paganin2002} (Fig.~4(c)), reflected in the difference image, Fig.~\ref{fig:Experimental}(e). In particular, the Laplacian-type character of this difference map is a signature of the higher-resolution projected thickness reconstruction associated with the Fokker--Planck analysis \cite{GPM2020}. This difference originates from the fact that the reconstruction using our new Fokker-Planck method properly separates phase from dark-field effects, while the TIE-based method interprets decreased-visibility phase contrast fringes as more-slowly-changing sample thickness. This higher spatial resolution associated with the Fokker-Planck analysis is more clearly seen with the seed pod sample presented in Appendix IV. 

With the thickness reconstruction using our method in hand, we calculate the last two terms in the dark-field retrieval, representing the estimated intensity image in the absence of dark-field, shown in Fig.~\ref{fig:Experimental}(f). The visibility reduction effects described in Fig.~\ref{fig:profiles} can be seen in the experimental data in Fig.~\ref{fig:Experimental}(i). This shows the profiles taken across the observed image (blue, Fig.~\ref{fig:Experimental}(b), first term on the second line of (\ref{eqn:Dlap})) and the image that would be expected in the absence of dark-field (red, Fig.~\ref{fig:Experimental}(f), second term on the second line of (\ref{eqn:Dlap})), with a difference in visibility in the profiles in regions where there is a strong dark-field signal, but a closer match between the profiles in regions where there is a weaker dark-field signal. By taking the difference between Fig.~\ref{fig:Experimental}(b) and (f), according to the bottom line of (\ref{eqn:Dlap}), we obtain panel (g), which highlights these differences in visibility between the blue and red profiles. Taking the inverse Laplacian, we obtain panel (h) which shows the recovered dark-field signal as described by (\ref{eqn:DF}). It is evident from panel (h) that there is virtually no scattering from the region outside the polystyrene microspheres, as expected. Of particular interest are the regions where the liquid the microspheres came in had not fully evaporated and as a result, the microspheres are clumped together in a wet mass and hence there are fewer air/plastic interfaces than the surrounding regions, and hence weaker dark-field. These regions within the sample are seen as regions of increased thickness in the retrieved thickness image (panel (d)) and regions of reduced scattering in the recovered dark-field signal (panel (h)). These observations help to demonstrate the complementary nature of the sample thickness/phase and dark-field in terms of the sample information these signals provide.  

\section{Discussion}

This paper presents a novel algorithm for thickness and dark-field retrieval from multiple-distance propagation-based x-ray imaging, via the Fokker--Planck equation. The simulation results show that this algorithm can extract dark-field effects that are independent of the sample phase and attenuation effects. The successful application of our approach to experimental data was also shown. To our knowledge, the use of propagation-based phase-contrast imaging to extract dark-field signals has only been demonstrated in one other paper, by Gureyev \textit{et al.}\cite{Gureyev2020}, where an alternative approach was taken, based on linearizing the Fresnel integral in the near-field regime. The method presented there has the advantage of only needing one intensity image to reconstruct both the projected thickness and dark-field signal of a homogeneous sample. As such, there is no potential complication due to misalignment between images taken at different propagation distances. However, we speculate that this use of a solitary intensity image may come at some cost, for example limiting how quantitative the reconstruction can be. Additionally, a two-image approach such as our method may help with separating phase and dark-field effects with high signal-to-noise ratio (SNR) and high spatial resolution -- at small propagation distances, a single-image approach may be less sensitive to dark-field, while at high propagation distances, dark-field blurring may reduce the spatial resolution of the resulting images retrieved with a single-image approach, as phase effects will be blurred out and it will not be clear whether the sample is slowly-varying or is producing a strong dark-field signal. A two-image approach avoids both issues. Our method extracts the dark-field signal as defined by the x-ray Fokker--Planck equation, so our method is sensitive to any local blurring, whether that be bulk scattering from unresolved microstructure or edge scattering. 

Upon taking the $D\qty(x,y)\rightarrow 0$ limit of the dark-field retrieval equation, (\ref{eqn:DF}), we recover the homogeneous phase retrieval method of Paganin \textit{et al.}\cite{Paganin2002}, which is based on the TIE. This result is not unexpected, since the Fokker--Planck equation, upon which our method is based, is the natural diffusive generalization of the TIE and hence, in the limit of a vanishing dark-field signal, the Fokker--Planck equation reduces to the TIE (as can be seen from (\ref{eqn:FPE})). While the TIE phase retrieval method of \cite{Paganin2002}, requires only one image, the method presented here uses this second measurement to extract the dark-field signal, which contains complementary information about the sample relative to its projected thickness. This separation of the projected thickness and dark-field signals serves the additional purpose of ensuring that the presence of dark-field effects does not result in inaccuracies in the projected thickness reconstruction. The difference between our method and the TIE-based method of Paganin \textit{et al.}~may be summarized as follows. In the latter method, the projected thickness is retrieved from contrast generated by attenuation, phase, and dark-field effects, while in our method the projected thickness is ideally only sourced from attenuation and phase effects, since the dark-field effects have been disentangled, with the added advantage of obtaining a complementary dark-field signal. This recovered dark-field signal can be useful for qualitative inspection and  visualizing/discerning various features. The recovered dark-field signal may also be converted into other useful quantities, such as the divergence angle of the SAXS cone, the blur width associated with the SAXS cone and the characteristic transverse length scale of the rapid spatial wave-front fluctuations induced by the unresolved microstructure present in the sample (see Fig.~3 in \cite{Paganin2019} for details). Note that wavefield discontinuities introduced by sharp edges present in the sample may also contribute to the recovered dark-field signal, since an edge can create a boundary wave that reduces local contrast (e.g. see Fig.~2(b) in Groenendijk \textit{et al.}\cite{Groenendijk}) and can also create a propagation-based fringe that increases local contrast\cite{Morgan2019}.
\par
In using the method presented here, it is advisable to include a small propagation distance that satisfies the near-field condition when performing the phase retrieval, so that high-spatial-frequency sample features can be resolved, since these features can be blurred out by dark-field effects at larger propagation distances (see the video included as Supplementary Materials II). It is then advisable to include a longer propagation distance, where dark-field effects are visually apparent, for the dark-field retrieval, noting these should still satisfy the near-field condition (\ref{eqn:Fresnel}). The optimal propagation distances for our method are highly sample-dependent, so optimization may be best achieved by looking for some visible blurring in the captured experimental propagation-based images, then choosing distances approaching this. It is also beneficial for there to be contrast across the whole field of view in the experimental images captured for analysis with our retrieval method, since an area of no local contrast would not be able to reveal any local blurring effects due to dark-field. As shown in Fig.~\ref{fig:profiles}, this can be either attenuation or phase contrast. In terms of spatial coherence, our algorithm requires that the characteristic length scale of source-size blurring, $w_{\text{source}}$, be less than the characteristic length scale of diffusive contrast due to SAXS, $w_{\text{SAXS}}$, so that the SAXS-associated blurring by the sample can be detected. That is, $w_{\text{source}}<w_{\text{SAXS}}$, which can be written as $s<R\theta$, where $s$ is the source size, $R$ is the source-to-sample distance, and $\theta$ is the opening angle of the SAXS cone. 
  
Compared to existing methods for extracting x-ray dark-field signals, such as analyzer-based imaging or grating interferometry, the method presented here does not require the careful alignment and stability that comes with using specialized optics. Additionally, since propagation-based phase-contrast imaging has been shown to be robust with respect to the use of polychromatic radiation\cite{Wilkins1996} and can account for finite source sizes\cite{Beltran2018}, we expect that our method will also be robust to the use of polychromatic radiation and transferable to lower coherence sources, such as medical or laboratory settings, pending future investigation. Note that the effects of polychromatic radiation could potentially be incorporated by multiplying (\ref{eqn:FPE}) with a wavelength-dependent weighting factor that describes the polychromatic x-ray spectrum and then integrating over the wavelength. However, since dark-field effects must be visualized directly, smaller pixels will likely be required in order to use our retrieval method with samples which scatter weakly. As a consequence, this may affect the size of samples that can be imaged on a given detector. Compared to grating-based methods for extracting dark-field signals, our method provides a dark-field reconstruction with less high frequency noise. This can be seen, for instance, by comparing the dark-field signal obtained from the sample shown in section IV using our Fokker-Planck retrieval method (Fig.~\ref{fig:Experimental}(h)), to the dark-field signal/scattering angle of a very similar sample (composed of the same material) obtained using single-grid imaging, as shown in Fig.~4 of \cite{How}. In addition, our method is also computationally fast and deterministic compared to the cross-correlation method used in \cite{How}. However, only a single sample exposure is required in the single-grid imaging technique, meaning the sample receives a lower radiation dose if a single exposure is used, compared to our Fokker-Planck method, which requires two sample exposures to extract the sample thickness and dark-field signal. Additionally, due to the division-by-zero error at the origin of Fourier space in (\ref{eqn:invlap}) and the subsequent use of (\ref{eqn:replace}), our method may potentially give rise to more low frequency artifacts. 
\par
In addition to the advantages stated above, some limitations of our method which we can foresee include the following. Firstly, the use of two intensity images taken at different propagation distances comes with the potential for misalignment artifacts. Such artifacts, however, can be mitigated with the use of translation and registration software prior to the application of our method (as mentioned in section IV). Secondly, our method requires a slightly increased radiation dose to be delivered to the sample relative to conventional PBI phase retrieval\cite{Paganin2002} or single-exposure PBI dark-field\cite{Gureyev2020} as a result of using two intensity images compared to one. This increased radiation dose is balanced by the benefit of quantitatively extracting the sample thickness free from dark-field effects, as well as the dark-field signal, providing increased sample structure information. Note also that grating and crystal-based methods require sufficient exposures (e.g. seven) to sample a stepping or rocking curve in order to extract a dark-field image. Lastly, our method also requires there to be contrast across the whole field of view, as areas of no local contrast in a captured intensity image are unable to reveal any local blurring effects due to sample dark-field effects. Note that if the sample alone is so smooth as to not produce much local contrast, contrast could be introduced via means of a reference pattern without the need for alignment, for example with a patterned sample holder or with a textured garment for a patient to wear. 
\par
This introduction of contrast leads to a point of comparison between (i) the method developed in the present paper and (ii) imaging approaches that employ an optical element to extract x-ray dark-field signals.  Consider, for example, the methods of x-ray speckle tracking \cite{berujon2012,morgan2012, zdora} or single-grid x-ray imaging \cite{wen2010, morgan2013}.  In these methods, a spatially-random speckle membrane (e.g.~a piece of sandpaper) or a phase-shifting or attenuating grid is introduced to create a spatially rapidly-varying intensity reference pattern. Information regarding the absorption, phase shift, and dark-field signal associated with a sample can be inferred by looking at sample-induced changes in the local intensity, the transverse position of the reference features, and the visibility of those features, respectively, in these contexts.  In the method developed in our paper, {\em the object is self-referencing}. The sample itself is considered to create the reference pattern that is subsequently diffused upon free-space propagation, thereby enabling the dark-field signal to be quantitatively extracted from bright-field data.  Stated differently, the sample plays two roles in our method, namely the obvious role as an unknown object whose structure is to be retrieved, and additionally, the role of a highly-structured mask employed to elucidate certain properties regarding the sample.  The sample contains its own mask---creating its own reference speckle pattern, so to speak---since the resolved Fourier components of the sample microstructure constitute an `internal speckle grating' whose post-sample diffusion allows information regarding the spatially-unresolved microstructure to be extracted. In our method, no separate `reference' grid-only or speckle-only image is required, since the sample is the reference.
\section{Avenues for future research}
We provide three possible directions for future research, which build directly off the work presented in this paper, and will be the subject of future papers. The first of these directions is a generalization of our retrieval method to multi-material samples, where there are three key quantities to be reconstructed separately; the attenuation, the phase shift induced by the sample, and the dark-field signal. A reconstruction of these quantities could be achieved in a manner similar to that shown in this paper, but using three different propagation distances, where the exit-surface intensity and exit-surface phase of the x-ray wavefield do not need to be coupled, as compared to the case of a homogeneous sample\cite{Cloetens}. The use of three distinct propagation distances would allow one to decouple the phase and diffusion terms in (\ref{eqn:FPE}), and hence solve for the phase, attenuation, and dark-field signal. 
\par
The second possible direction for future research is an extension of our retrieval method presented in this paper to computed tomography (CT), thereby providing a method to perform dark-field CT using a propagation-based imaging setup. This could be achieved simply by acquiring a set of experimental projections, taken at different angles around the sample, at a minimum of two different propagation distances, so that our method described in this paper can be used to reconstruct the sample thickness and dark-field signal for each projection. These retrieved thickness and dark-field projections could then be combined into a 3D mapping of both sample density and the dark-field signal by utilizing standard CT reconstruction methods, such as filtered-back projection. 
\par
In addition to the two directions outlined above, our algorithm could also be extended to the case of `directional dark-field', where the transverse cross-section of the local SAXS cone is considered to be elliptical rather than rotationally symmetric. In such scenarios, the single diffusion coefficient $D\qty(x,y)$ may be replaced by a symmetric rank-two diffusion tensor\cite{Paganin2019}:
\begin{equation}
\label{eqn:Dtensor}
\small
D\qty(x,y)\rightarrow 
\mqty[D_{xx}\qty(x,y) & \tfrac{1}{2}D_{xy}\qty(x,y) \\ \tfrac{1}{2}D_{xy}\qty(x,y) & D_{yy}\qty(x,y)].
\end{equation}
With this modification, (\ref{eqn:FPE}) becomes:
\begin{align}
\label{eqn:directionalFPE}
I\qty(x,y,z=\Delta)&=I\qty(x,y,z=0)\\ \nonumber -&\frac{\Delta}{k}\grad_{\perp}\vdot\qty[I\qty(x,y,z)\grad_{\perp}\phi\qty(x,y,z)]_{z=0}\\ \nonumber
+&\Delta^{2}\pdv[2]{x}\qty[D_{xx}\qty(x,y)I\qty(x,y,z)]_{z=0}\\ \nonumber
+&\Delta^{2}\pdv{}{x}{ y}\qty[D_{xy}\qty(x,y)I\qty(x,y,z)]_{z=0}\\ \nonumber
+&\Delta^{2}\pdv[2]{y}\qty[D_{yy}\qty(x,y)I\qty(x,y,z)]_{z=0}.
\end{align}

\par
In closing, this manuscript describes a retrieval method which can quantitatively recover x-ray phase and dark-field signals without using optics. The method provided generalizes the TIE single-material phase retrieval algorithm of Paganin \textit{et al.}\cite{Paganin2002}, which has been widely adopted throughout the x-ray imaging community. Hence, we hope that our method may also be of use in a range of imaging problems. Our algorithm could also be retrospectively applied to existing multiple-distance data. Potential applications of our dark-field method include quantitative measurements of the air-sacs in the lungs \cite{kitchen2020, willer2021} or capturing industrial processes involving microstructure \cite{prade2015}. 
\section*{Appendix I - Computational recipe for the simulated dataset}
Here we provide a computational recipe to simulate propagation-based images that include dark-field effects, taken within the manuscript as inputs to our phase and dark-field retrieval algorithm in Section III. Given the complex refractive index of a homogeneous sample, where $\delta$ and $\beta$ are taken to be constant, and given the simulated thickness of the sample $T_{\text{sim}}\qty(x,y)$, we create the exit-surface wavefield
\begin{equation}
\label{eqn:Exitfield}
\psi\qty(x,y,z=0)=\exp\qty[-k\qty(\beta+i\delta)T_{\text{sim}}\qty(x,y)].
\end{equation}
We then calculate the propagated wavefield by using the Fresnel propagator, $\mathcal{D}_{\Delta}^{(\text{F})}$ \cite{Coherent}:
\begin{align}
\label{eqn:Propfield}
\psi\qty(x,y,z=\Delta)\approx \mathcal{D}_{\Delta}^{(\text{F})} \psi\qty(x,y,z=0).
\end{align}
The intensity of the propagated wavefield, $I_{\text{P}}\qty(x,y,z=\Delta)$, is
\begin{equation}
\label{eqn:Perfectint}
I_{\text{P}}\qty(x,y,z=\Delta)=\abs{\psi\qty(x,y,z=\Delta)}^{2},
\end{equation}
resulting in propagation-based images that would be seen if dark-field effects were not present. In order to incorporate dark-field effects, the simulated blur width as described in Section III, $w\qty(x,y)$, is used to blur these propagation-based images through a position-dependent blurring kernel, which we take to be a two-dimensional Gaussian function with standard deviation $\sigma\qty(x,y)=\sqrt{2 w\qty(x,y)}\Delta$ (cf.~(9) in \cite{Morgan2019}):
\begin{align}
\label{eqn:Blurredint}
I_{\text{B}}\qty(x,y,z=\Delta)=\int_{-\infty}^{\infty} \int_{-\infty}^{\infty}I_{\text{P}}\qty(x',y',z=\Delta)\\
\times \nonumber\frac{e^{-\frac{(x-x')^{2}+(y-y')^{2}}{2\sigma^{2}\qty(x',y')}}}{2\pi\sigma^{2}\qty(x',y')} \,dx'\,dy'.
\end{align}
This produces propagation-based images that are locally blurred by the presence of dark-field effects. These are then further blurred by a two-dimensional Gaussian PSF of $\alpha=2$ pixel standard deviation to describe detector and source-size blurring:
\begin{align}
\label{eqn:PSFint}
I_{\text{PSF}}\qty(x,y,z=\Delta)=\int_{-\infty}^{\infty} \int_{-\infty}^{\infty}I_{\text{B}}\qty(x',y',z=\Delta)\\
\times \nonumber\frac{e^{-\frac{(x-x')^{2}+(y-y')^{2}}{2\alpha^{2}}}}{2\pi\alpha^{2}} \,dx'\,dy'.
\end{align}
The intensity images calculated at various propagation distances according to (\ref{eqn:PSFint}) were taken as inputs to the simultaneous phase and dark-field retrieval algorithm presented in the manuscript, as shown in Fig.~\ref{fig:Sim}(c). Additionally, the simulated dark-field signal shown in Fig.~\ref{fig:Sim}(b) is obtained from the standard deviation of the Gaussian used in (\ref{eqn:Blurredint}) as follows (cf. Fig~3(d) in \cite{Paganin2019}): 
\begin{equation}
\label{eqn:Dsim}
D_{\text{sim}}\qty(x,y)=\frac{\sigma^{2}\qty(x,y)}{\Delta^{2}}=2w\qty(x,y).
\end{equation}
Note that although the model given here is tailored to the context of a homogeneous sample and a rotationally invariant SAXS cone, this model may be readily adapted to simulate a multi-material sample and/or a directional SAXS cone. 
\section*{Appendix II - Quantitative accuracy with propagation distance}
Here we provide a metric for our simulated dataset, shown in Section III, which quantifies the accuracy of the TIE method\cite{Paganin2002} and our Fokker-Planck method, across multiple propagation distances. This metric, along with the reconstructed images, demonstrates the consistency and robustness of our retrieval method when using different propagation distances to reconstruct the sample thickness and dark-field signal.  

Define the following root-mean-square error (RMSE) metric to quantify the accuracy of the thickness reconstructions: 
\begin{equation}
\label{eqn:rmse}
T_\text{RMSE}=\sqrt{\frac{\iint \abs{T_{\text{retrieved}}-T_{\text{simulated}}}^{2}dxdy}{\iint \abs{T_{\text{simulated}}}^{2}dxdy}},
\end{equation}
where the integrals are taken over the entire area of the images. Here, $T_{\text{retrieved}}$ is the retrieved sample thickness using either the method of \cite{Paganin2002} or our Fokker-Planck method, and $T_{\text{simulated}}$ is the simulated thickness shown in Fig.~\ref{fig:Sim}(a). From Fig.~\ref{fig:Thickrange} below, it is evident that regardless of which pair of propagation distances we use (from those listed in section III) to reconstruct the sample thickness according to (\ref{eqn:thickness}), we obtain an accurate reconstruction of the simulated thickness. Furthermore, based on the values given by (\ref{eqn:rmse}) (listed in the caption of Fig.~\ref{fig:Thickrange}), the thickness reconstructions obtained using our Fokker-Planck method are consistently better than or at least as good as the corresponding TIE thickness reconstruction. 
\begin{figure}[ht]
\centering 
\hspace*{-2em}
\includegraphics[scale=0.5]{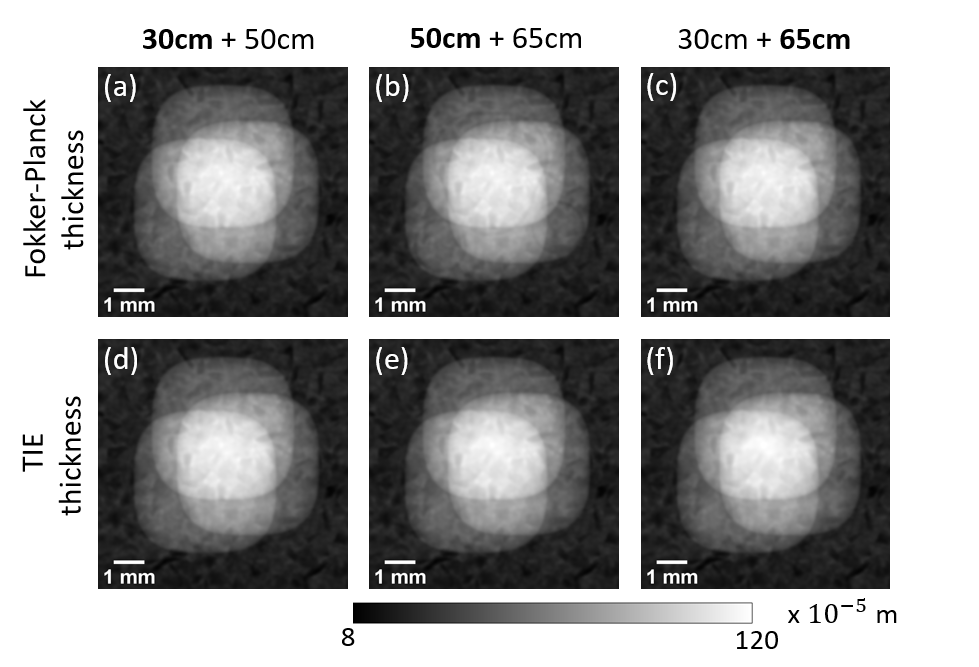}
\vspace*{0.3em}
\caption{Top row: Fokker-Planck thickness reconstructions using propagation distances of (a) \SI{30}{\centi\meter} and \SI{50}{\centi\meter} ($T_\text{RMSE}=0.016$), (b) \SI{50}{\centi\meter} and \SI{65}{\centi\meter} ($T_\text{RMSE}=0.017$), and (c) \SI{30}{\centi\meter} and \SI{65}{\centi\meter} ($T_\text{RMSE}=0.015$). Bottom row: TIE thickness reconstructions \protect\cite{Paganin2002} using propagation distances of (d) \SI{30}{\centi\meter} ($T_\text{RMSE}=0.016$), (e) \SI{50}{\centi\meter} ($\text{RMSE}=0.019$), and (f) \SI{65}{\centi\meter} ($T_\text{RMSE}=0.023$), corresponding to the bolded distances.}
\label{fig:Thickrange}
\end{figure}
Analogous to (\ref{eqn:rmse}), we can define a metric which quantifies the accuracy of the dark-field reconstructions using our method compared to the simulated dark-field signal shown in Fig.~\ref{fig:Sim}(b), which we call $DF_{\text{RMSE}}$:
\begin{equation}
\label{eqn:dfrmse}
DF_\text{RMSE}=\sqrt{\frac{\iint \abs{D_{\text{retrieved}}-D_{\text{simulated}}}^{2}dxdy}{\iint \abs{D_{\text{simulated}}}^{2}dxdy}}.
\end{equation}
Here, $D_{\text{retrieved}}$ is the reconstructed dark-field signal according to (\ref{eqn:DF}), $D_{\text{simulated}}$ is the simulated dark-field signal, and the integrals are taken over the whole area of the images, as above. We have scaled the simulated dark-field signal (see Fig.~\ref{fig:Sim}(b)) to match the range of values present in the retrieved dark-field signal, i.e. to span the grayscale bar shown in Fig.~\ref{fig:DFrange}.
\begin{figure}[ht]
    \centering
    \hspace*{-1.1em}
    \includegraphics[width=\linewidth]{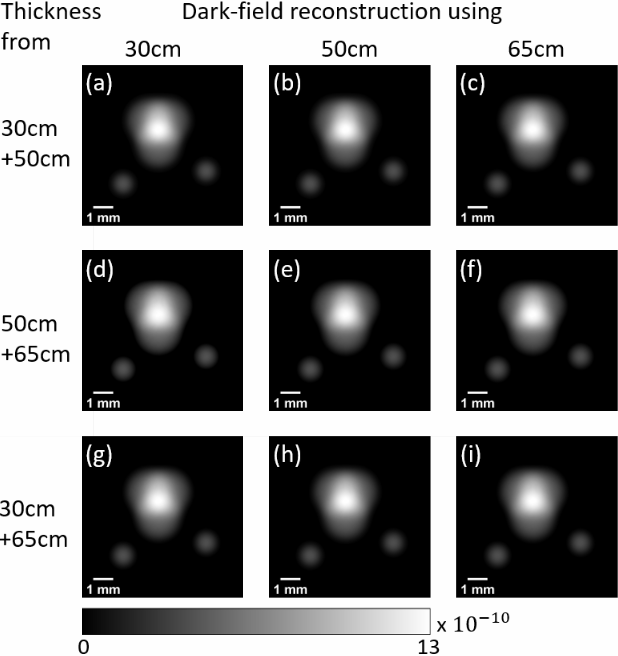}
    \caption{Various dark-field reconstructions at the different propagation distances of \SI{30}{\centi\meter}, \SI{50}{\centi\meter} and \SI{65}{\centi\meter}, along with the corresponding Fokker-Planck thickness reconstructions depicted in the top row of Fig.~\ref{fig:Thickrange}. (a) $DF_{\text{RMSE}}=0.147$, (b) $DF_{\text{RMSE}}=0.147$, (c) $DF_{\text{RMSE}}=0.148$, (d) $DF_{\text{RMSE}}=0.150$, (e) $DF_{\text{RMSE}}=0.153$, (f) $DF_{\text{RMSE}}=0.153$, (g) $DF_{\text{RMSE}}=0.149$, (h) $DF_{\text{RMSE}}=0.149$, (i) $DF_{\text{RMSE}}=0.149$.}
    \label{fig:DFrange}
\end{figure}
As can be seen from Fig.~\ref{fig:DFrange}, there is no discernible difference between the reconstructed dark-field signals using different propagation distances, and each reconstruction provides an accurate representation of the simulated dark-field signal.
\section*{Appendix III - The effects of noise on retrieval}
To perform a preliminary investigation into the effects of noise on the retrieval process described in this paper, we incorporate noise into our simulations from section III, as shown in Fig.~\ref{fig:noise} below. Here, noise has been incorporated in the form of Poisson noise with a signal-to-noise ratio (SNR) given by $\sqrt{N}$, where $N$ is the average number of x-ray photons arriving at the detector per pixel. The simulated intensity data at a propagation distance of \SI{65}{\centi\meter}, along with the retrieved sample thickness using both the TIE\cite{Paganin2002} and our Fokker-Planck method, as well as the recovered dark-field signal are shown in Fig.~\ref{fig:noise} as a function of SNR ranging from a value of 1000 down to a value of 70, with a higher SNR indicating a lower relative noise level and hence a less noisy reconstruction of the sample thickness and dark-field signal. 
\begin{figure*}
\centering 
\includegraphics[scale=0.8]{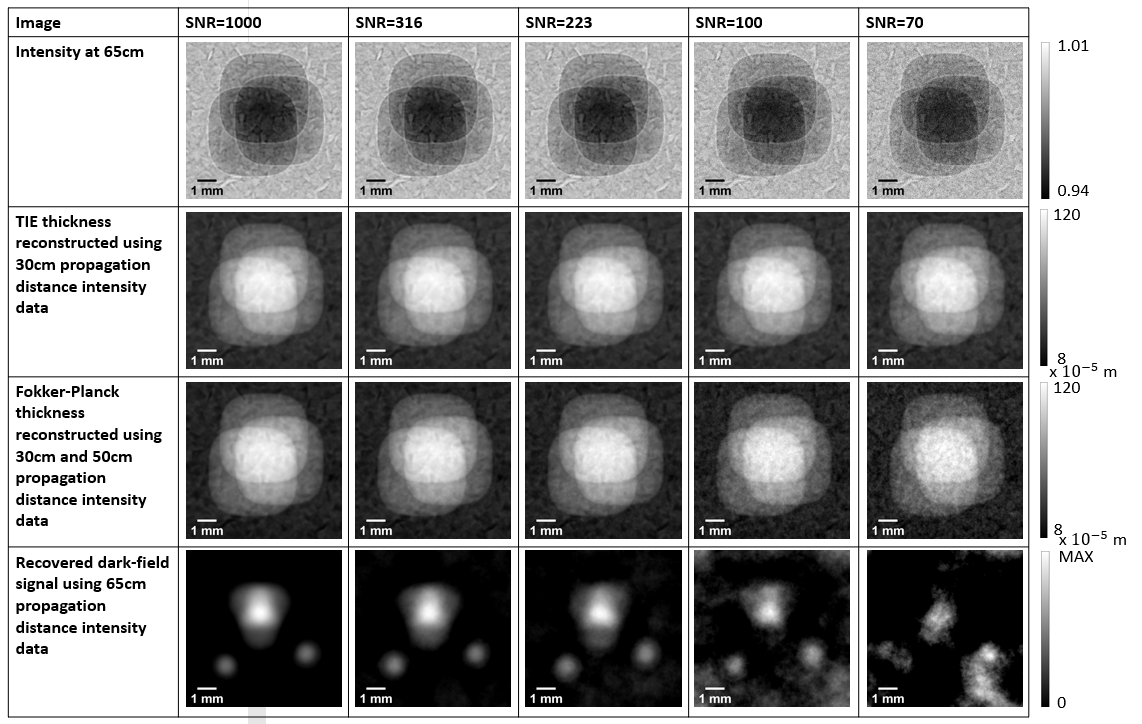}
\caption{Reconstructed sample thickness and dark-field signal using our method outlined in section II in simulation as a function of signal-to-noise ratio (SNR), with the SNR decreasing from left to right. Top row: Simulated intensity data at \SI{65}{\centi\meter}, Second row: Reconstructed sample thickness using the TIE based method of \protect\cite{Paganin2002}, Third row: Reconstructed sample thickness according to (\ref{eqn:thickness}) using our Fokker-Planck method, Bottom row: Recovered dark-field signal according to (\ref{eqn:DF}). From left to right, MAX$=8.9\times 10^{-10}, 9.2\times 10^{-10}, 9.4\times 10^{-10}, 9.8\times 10^{-10}, 6.8\times 10^{-10}$.}
\label{fig:noise}
\end{figure*}
It is worth noting here that the sensitivity of the retrieval and robustness to noise will likely depend upon the visibility of the intensity oscillations in the collected images and the strength of the dark-field signal. For example, an SNR value of 70 will not necessarily mean a retrieved sample thickness and dark-field signal as seen in the last column of Fig.~7 above. Hence, the SNR values quoted above should not be taken as defining values. 
\par
Two main observations can be drawn from Fig.~\ref{fig:noise}. The first of these is that the reconstruction process is successful, at least qualitatively in the case of the dark-field signal, for SNR values of 1000, 316 and 223. The second observation to be made is that the sample thickness retrieved using our Fokker-Planck method is not as robust to the presence of noise as the TIE method of Paganin \textit{et al.}\cite{Paganin2002}, noting that this method is well-known for strong noise-suppressing properties \cite{gureyev2017}. This is due to the use of two intensity images as compared to one in TIE phase retrieval, meaning there are two terms affected by the presence of noise in our new retrieval method. In particular, at SNR values of 100 and 70, the breakdown of the retrieved sample thickness using our Fokker-Planck method is evident compared to the corresponding TIE thickness reconstructions. This breakdown in the Fokker-Planck retrieved sample thickness is correlated with a significant degradation in the quality of the recovered dark-field signal, as can be seen from the bottom row of Fig.~\ref{fig:noise}. Hence, while the relative lack of robustness to noise in the thickness portion of our retrieval method as compared to TIE phase retrieval is a limitation of our method, one needs to keep in mind that the main benefit of our method is the ability to obtain a map of \textit{both} the sample thickness and dark-field signal corresponding to SAXS. At low SNR values, the recovered dark-field signal is poor and so the use of our method is of little to no value at such SNR levels. Conversely, at moderate and high SNR levels, the reconstruction of the sample thickness and dark-field signal provided by our method is accurate (and improves with higher SNR), and so at these types of SNR levels, our method has substantial benefit over TIE phase retrieval, as described at length in this paper. In practice, the amount of noise present in the retrieval process can be reduced by increasing the exposure time, by capturing multiple exposures at each propagation distance and then averaging the captured exposures or by pre-filtering of the images to remove high-frequency noise. 
\section*{Appendix IV - Retrieval from more complicated samples}
Here we test the limits of our retrieval method by considering a sample for which the $\delta/\mu$ ratio is not well known and which violates the assumption of azimuthally isotropic scattering. The purpose of doing this is to demonstrate that our new retrieval method can still be applied to samples that seemingly exceed the original domain of validity of our method, much in the same way the TIE retrieval method of \cite{Paganin2002} has been extended past its original domain of validity.  To this end, we captured propagation-based intensity images of a small plastic tube filled with agarose powder, attached to a green seed pod from a {\em Liquidambar styraciflua} tree using Kapton tape, on IMBL at the Australian Synchrotron using \SI{25}{\kilo\electronvolt} x-rays. By approximating the sample to have the refractive properties of PMMA, we estimated $\delta=\num{4.27e-7}$. Further, by measuring the attenuation relative to the sample thickness, we estimated the linear attenuation coefficient of the sample to be $\mu=\SI{107}{\meter^{-1}}$. The detector pixel size was \SI{10}{\micro\meter} and the images were captured at propagation distances of \SI{0.5}{\meter}, \SI{1}{\meter}, \SI{2}{\meter}, \SI{3}{\meter}, \SI{4}{\meter}, \SI{5}{\meter}, \SI{6}{\meter} and \SI{7}{\meter}. The retrieval results using the images taken at \SI{2}{\meter} and \SI{3}{\meter} are shown in Fig.~\ref{fig:Seedpod} below. The full sequence of collected images can be found in Supplementary Material II, where the reduction in local contrast across the tube of agarose powder, with increasing propagation distance, should be noted. Prior to applying our retrieval algorithm, all images were aligned and demagnified as outlined in section IV. 
\begin{figure*}[ht]
    \centering
    \includegraphics[scale=0.9]{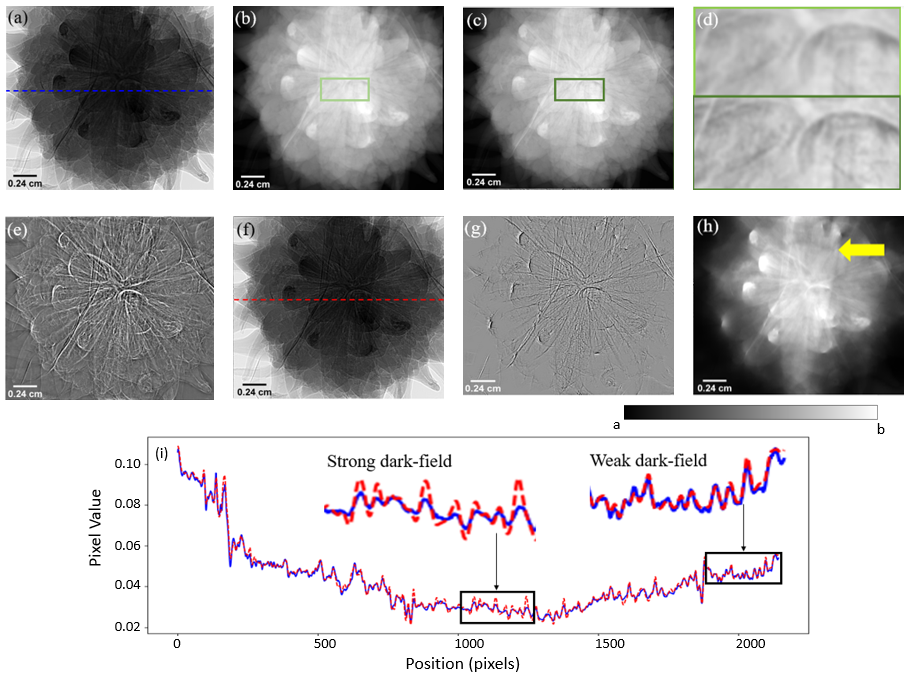}
    \caption{(a) Scaled propagation-based intensity image (first term in the last line of (\ref{eqn:Dlap})) of a \textit{Liquidambar styraciflua} seed pod with a plastic tube of agarose powder attached, taken with \SI{3}{\meter} propagation from sample to detector ($a=\SI{0.02}{\meter^{-2}}$, $b=\SI{0.15}{\meter^{-2}}$). (b) Retrieved thickness using TIE method with image taken at \SI{2}{\meter} propagation ($a=\SI{0}{\meter}$, $b=\SI{0.015}{\meter}$). (c) Retrieved thickness using our Fokker--Planck method with images taken at \SI{2}{\meter} and \SI{3}{\meter} propagation ($a=\SI{0}{\meter}$, $b=\SI{0.015}{\meter}$). (d) Color-coded magnified regions of the thickness maps in (b) and (c) respectively. (e) Difference between (b) and (c) ($a=\SI{-3e-4}{\meter}$, $b=\SI{5.4e-4}{\meter}$). (f) Dark-field-free intensity estimation at \SI{3}{\meter} using Fresnel propagation -- a more accurate model of the second term in the bottom line of (\ref{eqn:Dlap}) ($a=\SI{0.02}{\meter^{-2}}$, $b=\SI{0.15}{\meter^{-2}}$). (g) The Laplacian of the product of the dark-field signal and the $\exp[-\mu T\qty(x,y)]$ term, corresponding to (\ref{eqn:Dlap}), showing only those values where the dark-field coefficient is non-negative (i.e.~visibility of panel (f) is greater than or equal to that of panel (a)) ($a=\SI{-0.05}{\meter^{-2}}$, $b=\SI{0.03}{\meter^{-2}}$). (h) Dark-field signal found using (\ref{eqn:DF}) ($a=0$, $b=\num{1.4e-9}$). The yellow arrow indicates the level to which the powder reaches within the tube. (i) Profiles taken along blue and red dashed lines in (a) and (f), showing differences in visibility between the two terms on the last line of (\ref{eqn:Dlap}).}
    \label{fig:Seedpod}
\end{figure*}
As can be seen from the retrieved thickness maps shown in Fig.~\ref{fig:Seedpod}(b) and (c), using the TIE approach and our new Fokker-Planck method respectively, there is a clearly discernible higher spatial resolution associated with the projected thickness reconstruction using a Fokker-Planck analysis, with this higher spatial resolution being particularly evident in the zoomed-in regions shown in Fig.~\ref{fig:Seedpod}(d). This observation is further reinforced by the Laplacian-type character seen in the difference between the TIE and Fokker-Planck reconstructions shown in Fig.~\ref{fig:Seedpod}(e), as the observed Laplacian-type character of the difference map is a a signature of the higher spatial resolution associated to the Fokker-Planck analysis\cite{GPM2020}. This difference can be attributed to the fact that the Fokker-Planck analysis properly separates phase from dark-field effects, and that the TIE method interprets decreased-visibility phase contrast fringes as more slowly-changing sample thickness. 
\par
Using the reconstructed Fokker-Planck thickness, the last two terms in the dark-field retrieval step can be calculated, representing the estimated intensity image in the absence of any dark-field effects, shown in Fig.~\ref{fig:Seedpod}(f). For this dataset, instead of calculating the latter two terms in (\ref{eqn:Dlap}), we opted to use the Fresnel propagator to propagate the exit-surface wavefield, calculated from the retrieved Fokker-Planck thickness, to a distance of \SI{3}{\meter}, to more accurately describe the PBI fringes from a range of sample feature sizes and hence a range of Fresnel numbers. This replacement is valid since the Fresnel propagator reduces to the latter two terms of (\ref{eqn:Dlap}) (up to scaling factors of the propagation distance) in the TIE regime limit (see pp. 324--326 of \cite{Coherent} for details). Taking the profiles along the blue and red dashed lines in Fig.~\ref{fig:Seedpod}(a) and (f) respectively, we obtain Fig.~\ref{fig:Seedpod}(i), from which it can be seen that there is a difference in visibility in the profiles in regions in which there is a strong dark-field signal, while there is a closer match between the profiles in regions where there is a weaker dark-field signal. The next step in the dark-field retrieval process would be to take the difference between Fig.~\ref{fig:Seedpod}(a) and (f). Based on the profiles in Fig.~\ref{fig:Seedpod}(i), if we were to do this for this sample, where the dark-field is relatively weak, then we would see both regions where the experimental image has reduced local visibility, and occasionally regions where the experimental image has increased local visibility (e.g. see the strong PBI fringes around 170 pixels into the profile). By simply taking the difference between panel (a) and (f), we would lose information about where the experimental image has higher or lower visibility than we would expect for the given sample thickness in the absence of dark-field. However, since the dark-field signal is naturally restricted to be non-negative (see section I), any increase in local visibility of the experimental image is nonphysical (as this leads to a negative diffusion coefficient), with any such apparent increases in visibility being due to the presence of noise or slight mismatches in visibility from numerical modeling, and so we seek to isolate the non-negative dark-field signal here. To evaluate which profile has higher visibility, we measured the average curvature of the intensity oscillations in each direction by means of the absolute value of the second derivative of the intensity, which should oscillate between the value of the maximum curvature (e.g. peak/trough) and zero. To measure this average curvature in the presence of noise, we first utilized a two-dimensional Savitzky-Golay filter\cite{Press} to compute the Laplacian of the relevant images (Fig.~\ref{fig:Seedpod}(a) and (f)), a method that avoids noise amplification. We then took the absolute value of each of these Laplacian images and smoothed with a 2D Gaussian kernel of five pixel standard deviation in each direction to remove zeros. In image regions where there was an apparent increase in visibility (i.e. the blue profile had higher visibility than the red), Fig.~\ref{fig:Seedpod}(g) was set to zero. Finally, the dark-field signal was retrieved using (\ref{eqn:DF}) and is shown in Fig.~\ref{fig:Seedpod}(h). One can see that the agarose powder in the tube is more visible in Fig.~\ref{fig:Seedpod}(h), compared to the thickness reconstructions in (b) and (c), for instance by looking at the tip of the tube and the level to which the microstructures fill up the tube, indicated by the yellow arrow in panel (h). The internal wood-like structure of the seed pod also generates a strong dark-field signal, which is not seen in the thickness reconstructions. Note also that although the dark-field signal generated here by the seed pod is likely to be directional, the described dark-field retrieval algorithm is still able to reconstruct the dark-field image. In this case, we could expect the strength of the retrieved dark-field signal to correspond to the root mean square of the major and minor axis of directional dark-field.
\par
As a final point, we discuss the reason for using the additional filtering steps outlined in the previous paragraph when retrieving the dark-field signal from this sample as compared to the sample presented in section IV. As can be seen from Fig.~\ref{fig:Experimental}(i), there are very few locations in which the blue profile, corresponding to the scaled intensity image, is more visible than the red profile, which corresponds to the dark-field free intensity estimation. More importantly, the difference between the profiles in regions where this occurs (e.g. near the  label `weak dark-field') is much smaller in magnitude than the difference in regions where the red profile has a higher visibility than the blue (`strong dark-field' regions). By contrast, it can be observed from Fig.~\ref{fig:Seedpod}(i) that there are significantly more regions in which the blue profile has high visibility than the red profile, and in these regions, the difference between the blue and red profiles is comparable to the difference seen between the profiles in regions where the red profile has a higher visibility than the blue profile. It is precisely when these differences in magnitude become comparable where we would lose information about which profile has higher visibility by simply taking the difference between the profiles, and hence the additional steps in the dark-field retrieval process, described in the previous paragraph, become important. 
\bibliographystyle{IEEEtran}
\bibliography{references}

\end{document}